# Elastic Strain Fields In Lateral Double Ge/Si Quantum Dots

Yu. N. Morokov*, M. P. Fedoruk

*Institute of Computational Technologies SB RAS, Novosibirsk, 630090 Russia and*
*Novosibirsk State University, Novosibirsk, 630090, Russia*

**Abstract** Simulations of the elastic strain fields for double Ge/Si quantum dots located on the same wetting layer are carried out. The cluster approximation is used for the atomistic model based on the Keating potential. The spatial distributions of the strain energy density and electron potential energy are calculated using clusters containing atoms of 150 coordination spheres. It is shown that the used cluster boundary conditions are close to the periodic boundary conditions.

## I. INTRODUCTION

In work [1] it was shown that the development of multilayer structures with vertically stacked Ge/Si(001) quantum dots makes it possible to fabricate deep potential wells for electrons with vertical tunnel coupling. In the work [2] it was given more detailed description of results of our calculations for strain fields for single Ge/Si quantum dots. In this article we present the results of calculations for strain fields for lateral double Ge/Si quantum dots.

Essential role in the self-assembly of nanostructures during the epitaxial growth of germanium on crystalline silicon is played by inhomogeneous elastic strains in the system [3-5]. The strain also influence on the shape of the potential wells for electrons and holes in semiconductors. There are a lot of theoretical works on the strain distribution in quantum dot structures [5], including works on simulation the electronic structure and elastic strain distributions in self-assembled Ge/Si quantum dots [6-9].

In [2], in particular, it has been shown that there are 10 regions of increased strain, spatially localized in silicon, and hence 10 potential wells in which formation of the localized quantum states for electrons is possible. Noticeable overlapping is possible between the potential wells of the adjacent quantum dots at growing of arrays of such quantum dots. It can significantly change the parameters of these wells, and also to facilitate the tunnel transfer of a charge between adjacent quantum dots.

The two variants of the relative positioning of the interacting quantum dots are the most significant: a vertical positioning, when one quantum dot is located above the other in the direction of epitaxial growth (along the $z$ axis), and a lateral positioning, when two adjacent quantum dots are located at the same wetting layer. Both variants can be used in the design of the possible nanoelectronic circuits of the organization of interaction between these quantum dots.

In this paper, we consider the second of these cases - the lateral positioning of two quantum dots (in the direction of the $x$-axis). A significant extension of the system in one direction (along the $x$ axis) require from us in this case more detailed discussion of influence of boundary conditions that we set at the cluster boundary.

## II. APPROACH DESCRIPTION

We consider two quantum dots arranged on the wetting layer five atomic layers thick. The considered germanium quantum dots are {105} facetted pyramids with a square base and a height-to-base size ratio of 1 : 10. Consideration of the influence of the thickness of the germanium wetting layer has been carried out in [2].

We use here the same model as in [2]. Consideration is conducted using the cluster approximation.

We use the following boundary conditions for the clusters. For atoms of two outer coordination shells (which also include germanium atoms of the wetting layer), the $x$- and $y$-coordinates are fixed, but the $z$-coordinates (in the growth direction) are completely free for the relaxation of all atoms. It becomes possible due to a rupture of crystal silicon by germanium of the wetting layer infinite in the $x$ and $y$ directions.

The calculations were performed for clusters containing atoms of 150 coordination shells; the central cluster atom is considered as belonging to the zeroth coordination sphere. The cluster of 150 coordination shells contains 2840951 atoms. Two last coordination spheres contain 111756 atoms, and, respectively, only atoms of 148 coordination spheres (2 729 195 atoms) are completely free for the relaxation.

In this work, all pyramids have a base half-width of 55 atomic layers. It corresponds to the full width of the pyramid base of 14.93448 nm (measured between the centers of uttermost atoms of the pyramids).

As the first step for each structure, the system elastic energy (Keating potential [11]) is minimized using the conjugate gradient method. The numerical calculation is completed, when the change in the total cluster energy at one conjugate gradient step becomes smaller than the total cluster energy by 14 orders of magnitude. The accuracy limitation is caused by rounding errors when using double-precision numbers.

As the second step, we calculate the components of the strain tensor using the algorithm presented in [10, 2].

At the final step, we estimate the values of the effective electron potential energy for each atom of the cluster, using the effective mass approximation and the obtained diagonal elements of the strain tensor [10, 12, 2].

The corresponding parameters of the model and details of algorithms were given in [2].

For more informative display of energies in the colors we use nonlinear color scale. The concordance between energy values and the color palette is established in each figure. More often, the central section ($y = 0$) of the clusters will be presented in the subsequent figures. Cases of consideration of other, lateral sections ($z = $ const), will be separately mentioned.

## II. RESULTS

The results of our calculations for the single pyramid having a base half-width of 55 atomic layers have been presented in Section II.2 in [2].

Further we consider two identical pyramids located at one wetting layer. The centers of the pyramids are displaced from each other along the $x$ axis (Fig. 1). Below we present the results of calculations for four different distances between the centers of the pyramids: 110, 130, 170, and 190 atomic layers. That is, the distances between the centers of the nearest atoms of the pyramids are equals, respectively, 0, 20, 60 and 80 atomic layers. In the first case it means not touch of the pyramids, but the overlapping of a single layer of atoms of the pyramids.

# 1. Double Quantum Dots with the distances between the centers of the pyramids equal to 130 and 170 atomic layers

Figure 1 shows the distribution of the strain energy density for the central section ($y = 0$) of the cluster in the case where the distance between the centers of the pyramids equals to 130 atomic layers. The width of the wetting layer along the $x$ axis is equal to the width of the cluster. The wetting layer contains atoms of 301 atomic layers (along the $x$ axis) and its width is equal 40.7304 nm (between the centers of the uttermost atoms).

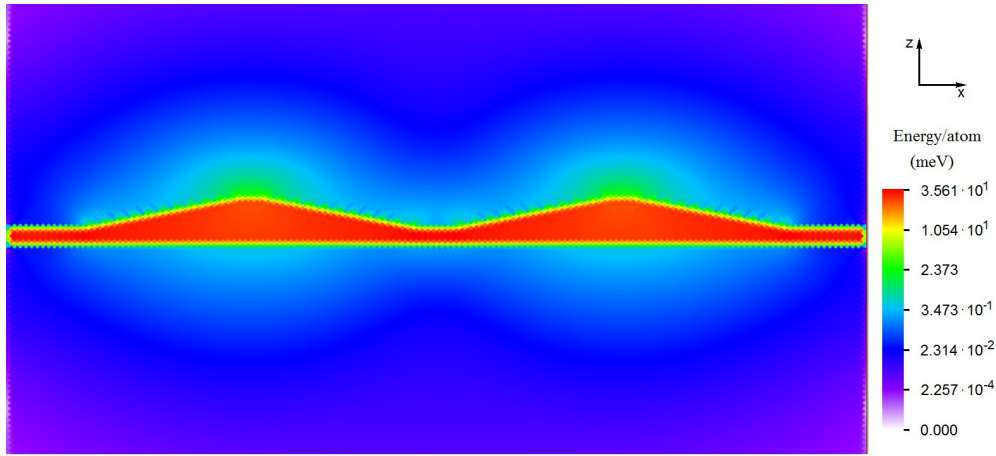

**Figure 1.** The distribution of the bulk density of the strain energy in silicon for two pyramids with the distance between the centers of the pyramids equals to 130 atomic layers.

The distribution of the electron potential energy in silicon for $\Delta^{001}$ valley in the central section of the same cluster is presented in Fig. 2. Below, by consideration of the electron potential energy, for example, for $\Delta^{100}$ valley we will automatically mean that this consideration applies also to $\Delta^{\bar{1}00}$ valley. Therefore valleys $\Delta^{\bar{1}00}$, $\Delta^{0\bar{1}0}$, and $\Delta^{00\bar{1}}$ will not be noted separately.

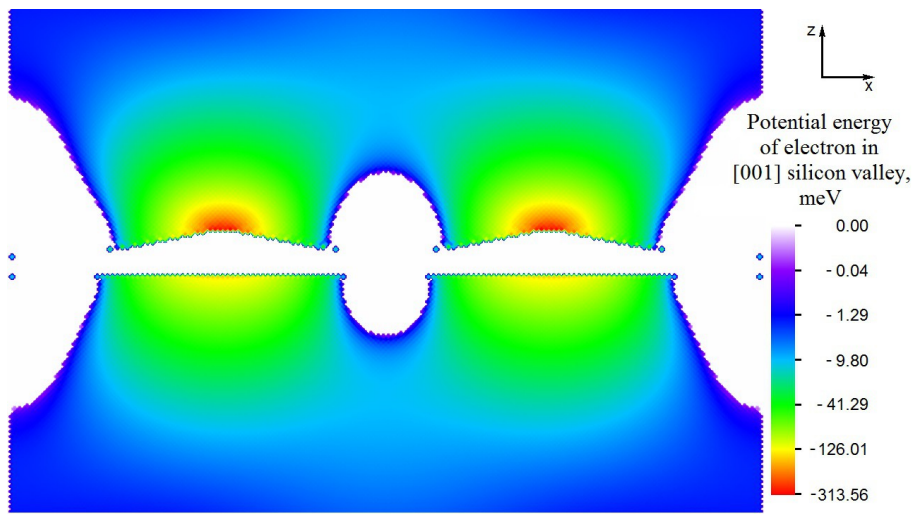

**Figure 2.** The distribution of the electron potential energy in silicon for $\Delta^{001}$ valley for two pyramids with the distance between the centers of the pyramids equals to 130 atomic layers.

We display in all figures only negative values of the electron potential energy in silicon (below the bottom of the conduction band of non-strained silicon), because we are interested in the possible localization of electrons in the quantum dot. The germanium atoms and the silicon atoms forming Si-Ge bonds are not displayed too.

It is visible from the Fig. 2 that the areas of negative values of electron potential energy for the $\Delta^{001}$ valley are overlapping above pyramids and below the wetting layer. At the same time the barriers of transition of an electron from one pyramid to another are reduced by only 9.6 meV and 10.0 meV, respectively. The potential wells become even smaller in comparison with single pyramids. Now the lowest value of the potential energy is equal to -316.37 meV (for the deeper wells above the tops of pyramids) while it was equal to -327.03 meV for a single pyramid. The depth of the potential wells under the pyramids also decreases. Now we obtain a value -147.01 meV instead of -151.71 meV for a single pyramid.

The last values of the calculated potential energy relate to individual atoms, without using interpolation between the nearest nodes, which we apply in case of graphic reproduction of the fields represented in the figures. Therefore, these values are slightly different from the values displayed on the palettes. So, exactly because of such interpolation the lower bound of the energy on the palette in Fig. 2 (-313.56 meV) differs from the extreme value reached on one of silicon atoms (-316.37 meV).

The situation is different for the valley $\Delta^{100}$, as can be seen from the following figures 3 and 4. Such close lateral layout of the pyramids leads to the overlapping of the potential wells and to deepening of their bottom from -109.18 meV (for wells above the wetting layer) to -129.27 meV. The deepening of the wells below the wetting layer is even more pronounced: from -46.77 meV to -72.43 meV.

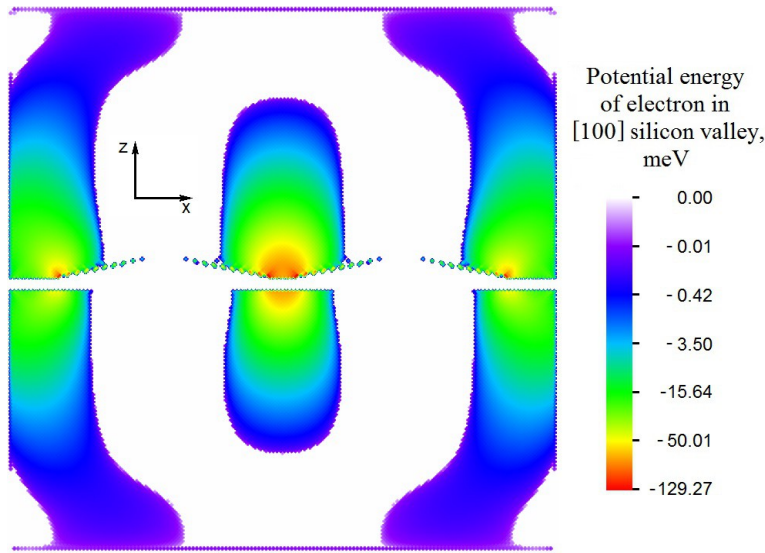

**Figure 3.** The distribution of the electron potential energy in silicon for $\Delta^{100}$ valley for two pyramids with the distance between the centers of the pyramids equals to 130 atomic layers.

There are barriers to possible transitions of an electron from one pyramid to another. Concerning the $\Delta^{100}$ valley we can talk about "horizontal" transition along *x*-axis between adjacent potential wells, located above the wetting layer, and between the potential wells below the wetting layer.

The potential wells merge in one limit case when the distance between the centers of the pyramids is less or equal to 109 atomic layers. The transition barriers are equal to zero in this case.

The electron transitions between the pyramids are also possible in the other limit case, at very large distances between the pyramids, but only through the conduction band. In this case, the barriers are simply equal to depths of the potential wells for a single pyramid, i.e., 109.18 meV and 46.74 meV, respectively, for the upper and lower potential wells.

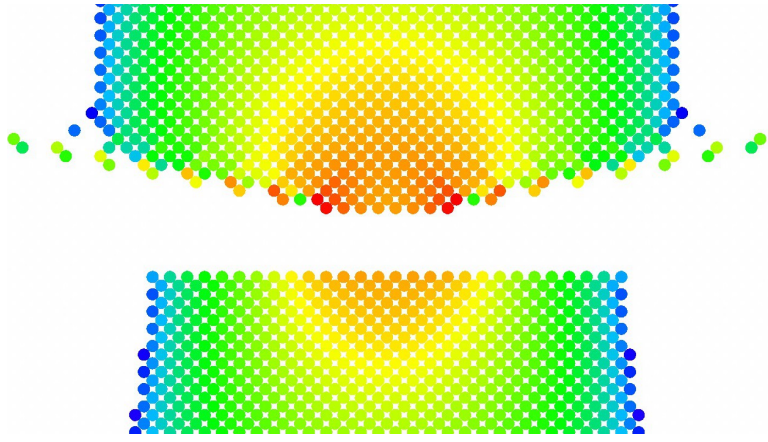

**Figure 4.** The enlarged central fragment of Fig. 3.

The barriers of transition gradually reduce to zero when the pyramids are approaching.

Of course, it is not about estimates of real quantum transitions, but simply about the analysis of the calculated effective potential surface within the model considered by us.

It is necessary for discussion of the possibility of quantum transitions, at least, to clarify the question: are the depth and the spatial extent of the potential wells enough for the localization of an electron in them or no? It requires the solution of the stationary Schrödinger equation for the calculated potential surfaces.

The two lower potential wells already are merged into one in the case shown in Figures 3 and 4, and, accordingly, there is no barrier for transition. The minimum of the obtained well, -72.43 meV, is reached in the middle, that is, at $x = 0$.

The upper wells are not yet merged, and their depth equals -129.27 meV. The saddle point between the wells is located at $x = 0$ higher along the axis z, at the level of silicon atoms of 7th layer over the wetting layer. The barrier between the upper wells is 55.36 meV when an electron passes through this point. This estimation for the barrier was performed already for the continuous potential surface displayed in Fig. 3. Here we already used the interpolation of potential energy between closest nodes for the section $y = 0$.

Influence of cluster boundaries along the $x$ axis for our two pyramids, in general, must become more noticeable for the $\Delta^{100}$ valley. X-coordinates of atoms of two last coordination spheres are frozen in our boundary conditions. If we fix $x$-coordinates only for atoms of the last coordination sphere on the left and right boundaries of the cluster, it actually would mean setting the periodic conditions along the $x$-axis for the components $u_{xx}$ of the strain tensor. However, we fix $x$-coordinates for atoms of the penultimate coordination sphere too. In this sense, our boundary conditions give some error in the solution of the periodic task.

The potential energy of an electron for the $\Delta^{100}$ valley is determined mainly just by the $u_{xx}$ component of the strain tensor [2]. Therefore, it is possible in a good approximation to present for Fig. 3, as well as for all other similar figures obtained by us, the periodic continuations to the left and to the right, outside the drawings. Periodic conditions for the Fig. 3 mean that there are other pyramids on the right and at the left, outside the drawing. The distances between the centers of these pyramids and the pyramids shown in Fig.3 are equal to 170 atomic layers.

Let's consider the results of the calculation of two pyramids, similar to the previous one, but now the distance between the centers of the pyramids is 170 atomic layers. The result for the $\Delta^{100}$ valley is presented in the Fig. 5. If in Fig. 3 the central and side potential wells corresponded to the distance between the centers of the pyramids 130 and 170 atomic layers, then in the Fig. 5 the situation is reversed - the corresponding distances are equal 170 and 130 atomic layers, respectively.

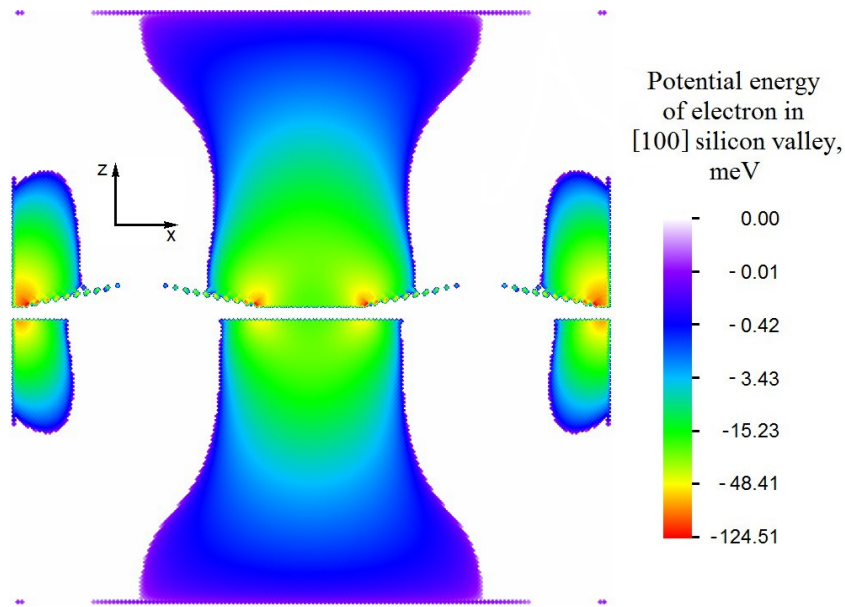

**Figure 5.** The distribution of the electron potential energy in silicon for $\Delta^{100}$ valley for two pyramids with the distance between the centers of the pyramids equals to 170 atomic layers.

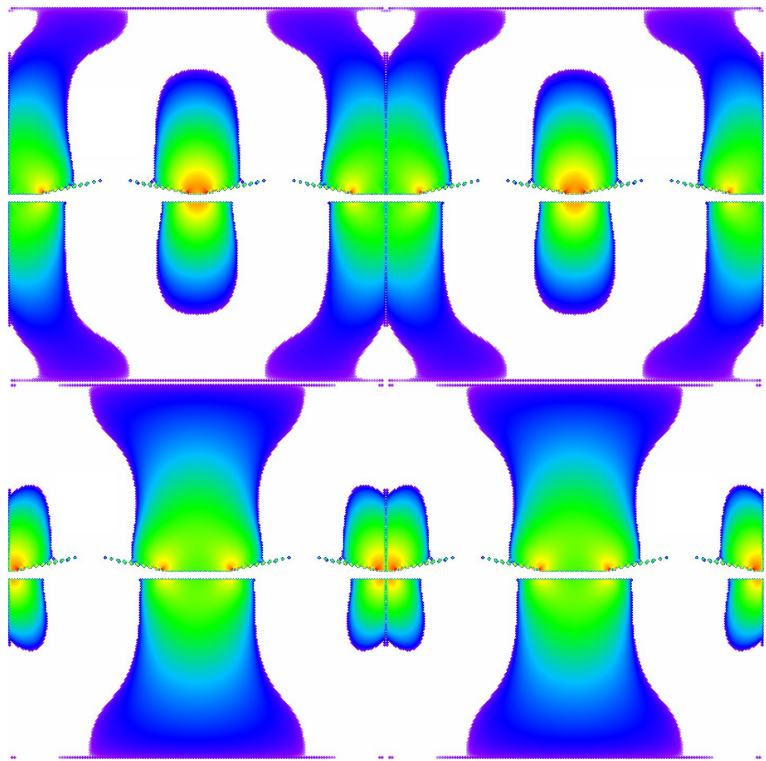

**Figure 6.** The distribution of the electron potential energy in silicon for $\Delta^{100}$ valley for the structure, periodic along the axis $x$, obtained by continuation of Fig. 3 and Fig. 5. The distances between the centers of the pyramids equal to 130 and 170 atomic layers.

Periodically continued (stuck together) images of Fig. 3 and Fig. 5 are shown in the Fig. 6 for comparison.

The section $z$ = const is presented in Fig. 7 and Fig. 8 for valleys $\Delta^{100}$ and $\Delta^{010}$.

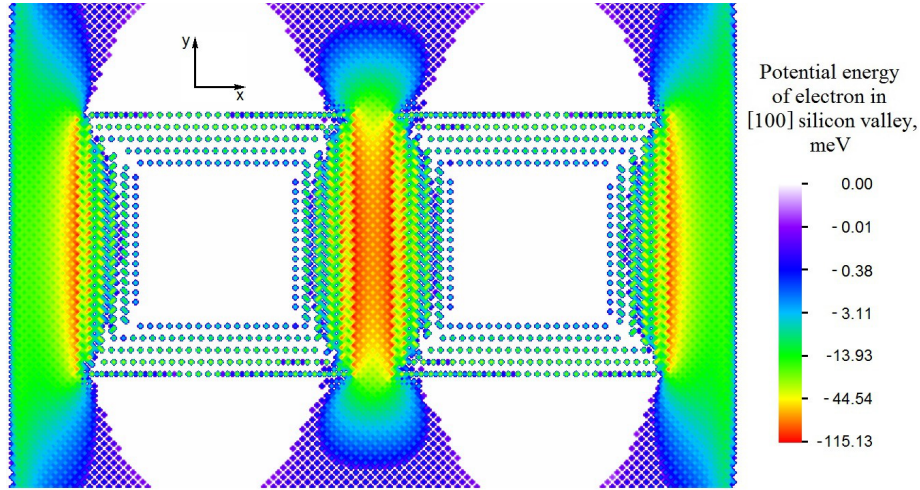

**Figure 7.** The distribution of the electron potential energy in silicon for $\Delta^{100}$ valley in the section $z$ = const (3rd atomic layer of pyramids) for two pyramids with the distance between the centers of the pyramids equals to 130 atomic layers.

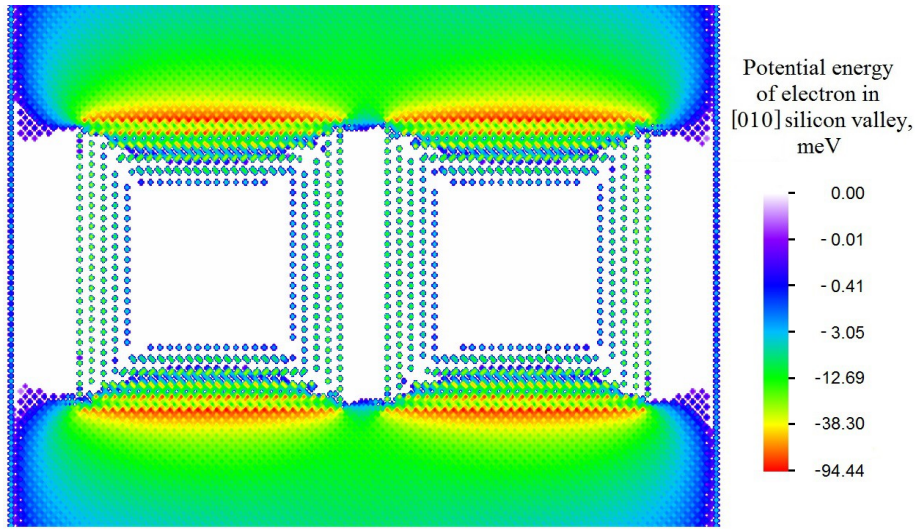

**Figure 8.** The distribution of the electron potential energy in silicon for $\Delta^{010}$ valley in the section $z$ = const (3rd atomic layer of pyramids) for two pyramids with the distance between the centers of the pyramids equals to 130 atomic layers.

## 2. Double Quantum Dots with the distances between the centers of the pyramids equal to 110 and 190 atomic layers

Let's reduce the distance between the centers of the pyramids - to the value of 110 atomic layers. Now, there are no silicon atoms between the pyramids in their zero layer. Fig. 9 shows the distribution of the strain energy density for the central section ($y$ = 0) for this case.

Reduction of the distance between the pyramids leads to even more noticeable reduction of the depth of the potential wells for an electron in the $\Delta^{001}$ valley (Fig. 10), up to -315.50 meV (upper) and -145.69 meV (lower). The barrier of the electron transition between the two upper wells is reduced by 18.91 meV, and for the transition between the lower wells by 20.07 meV in comparison with the electron transition to the conduction band.

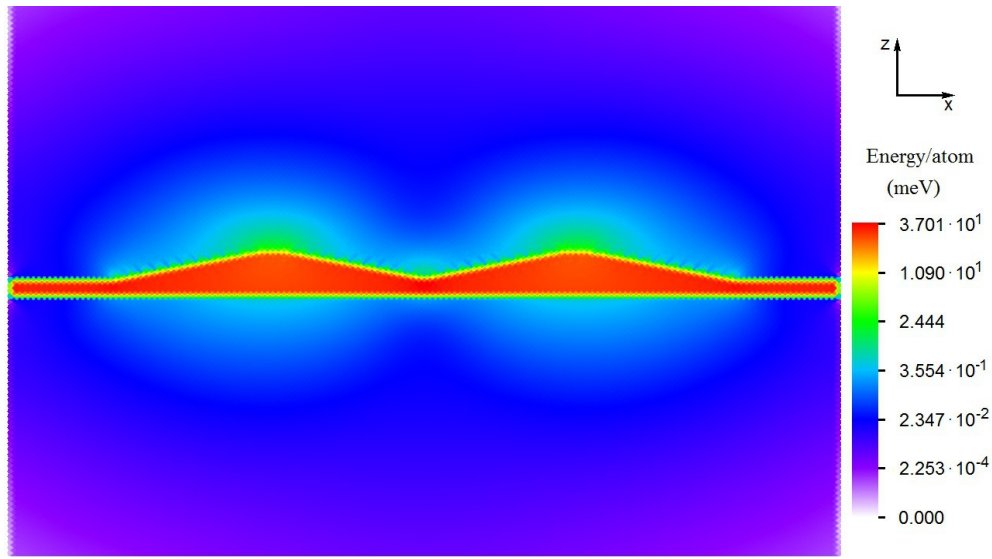

**Figure 9.** The distribution of the bulk density of the strain energy in silicon for two pyramids with the distance between the centers of the pyramids equals to 110 atomic layers.

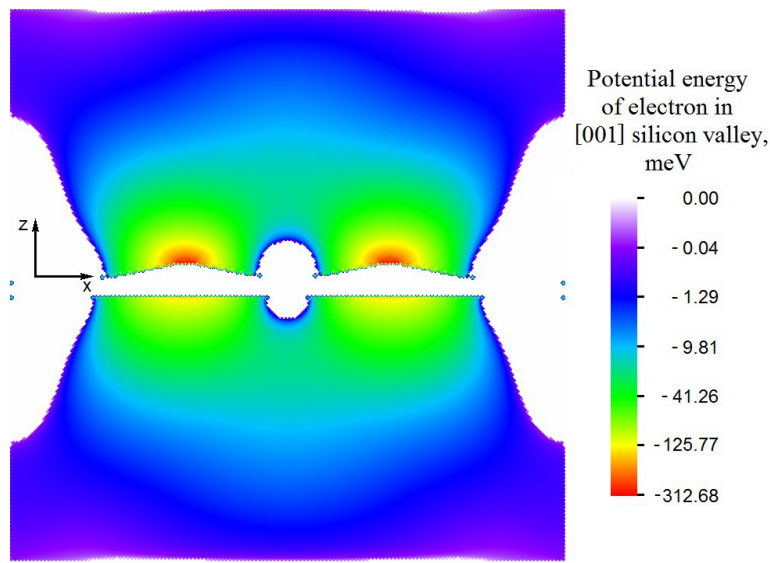

**Figure 10.** The distribution of the electron potential energy in silicon for $\Delta^{001}$ valley for two pyramids with the distance between the centers of the pyramids equals to 110 atomic layers.

To check the periodicity of the electron potential energy along the *x* axis, we involve the calculation of a cluster with two pyramids, the distance between the centers of which is equal to 190 atomic layers (Fig. 11). A specific of this structure is that two extreme layers of atoms of the pyramids are on the cluster boundary, and we don't carry out the relaxation of *x*-coordinates for these atoms. Therefore, the influence of the boundary conditions has to be the most essential here.

The periodic repeats for Fig. 10 and Fig. 11 are constructed in Fig. 12 by analogy with the Fig. 6.

The similar periodic construction (Fig. 15) is done below for the same two clusters for $\Delta^{100}$ valley just as it was made in Fig. 6. When building Fig. 15 we used the distributions of the electron potential energy given in Fig. 13 and Fig. 14.

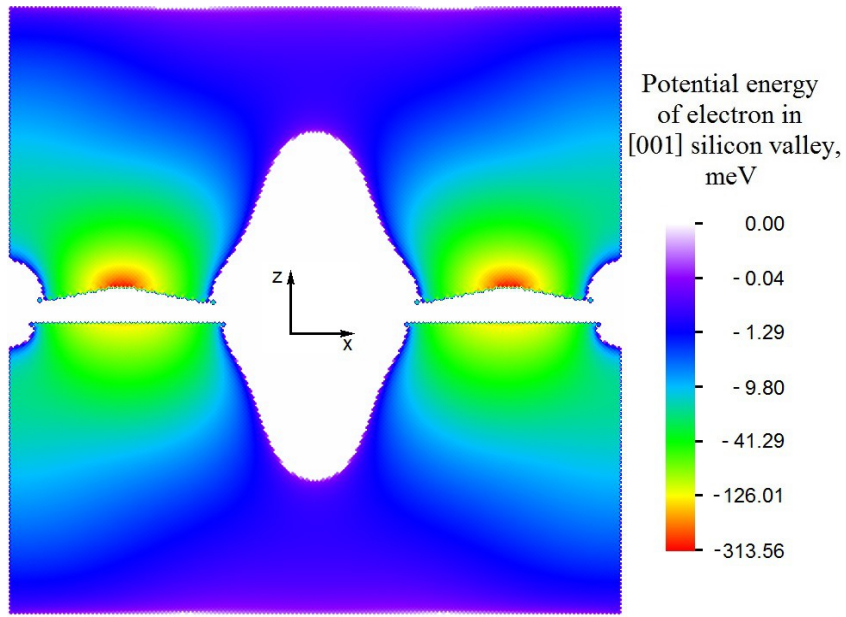

**Figure 11.** The distribution of the electron potential energy in silicon for $\Delta^{001}$ valley for two pyramids with the distance between the centers of the pyramids equals to 190 atomic layers.

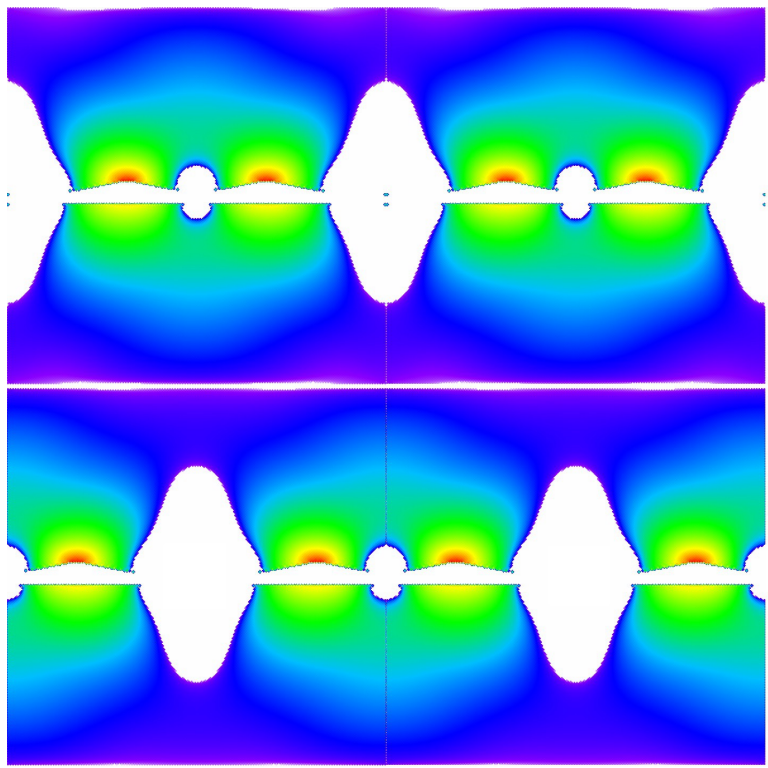

**Figure 12.** The distribution of the electron potential energy in silicon for $\Delta^{001}$ valley for the structure, periodic along the axis *x*, obtained by continuation of Fig. 10 and Fig. 11. The distances between the centers of the pyramids equal to 110 and 190 atomic layers.

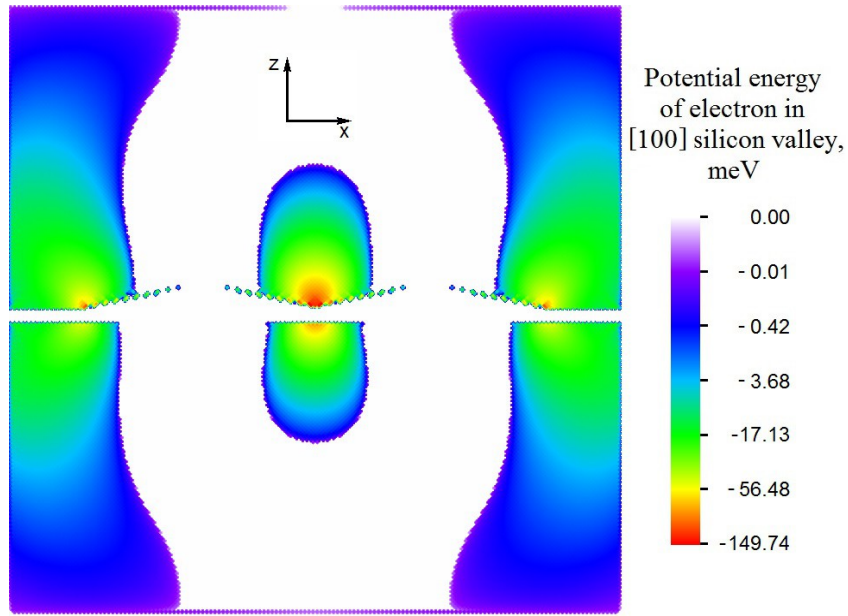

**Figure 13.** The distribution of the electron potential energy in silicon for $\Delta^{100}$ valley for two pyramids with the distance between the centers of the pyramids equals to 110 atomic layers.

For $\Delta^{100}$ valley, as can be seen from Fig. 13, the potential wells of the pyramids merge in the center of the cluster. They are deepened to energy values -149.76 meV (the upper well) and -82.73 meV (the lower well).

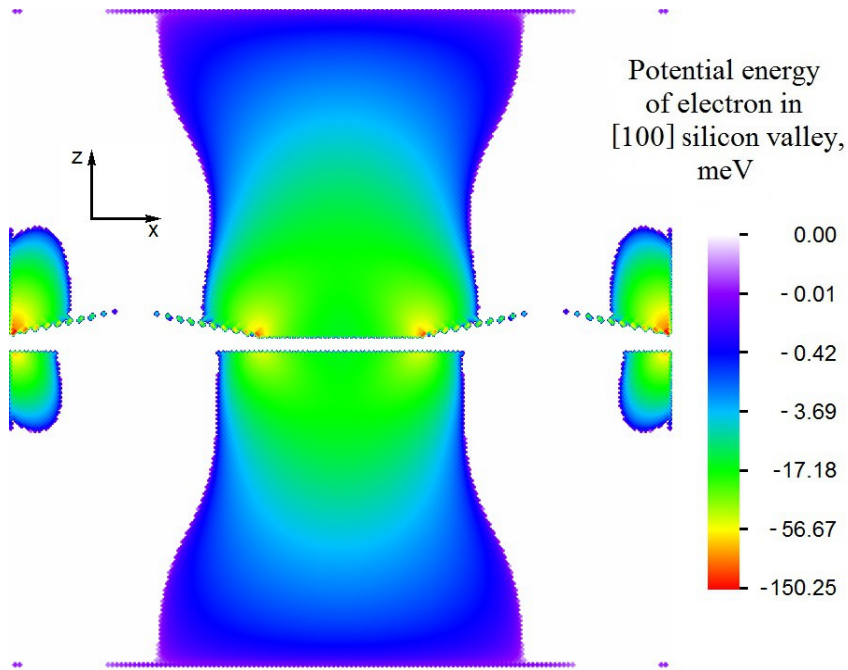

**Figure 14.** The distribution of the electron potential energy in silicon for $\Delta^{100}$ valley for two pyramids with the distance between the centers of the pyramids equals to 190 atomic layers.

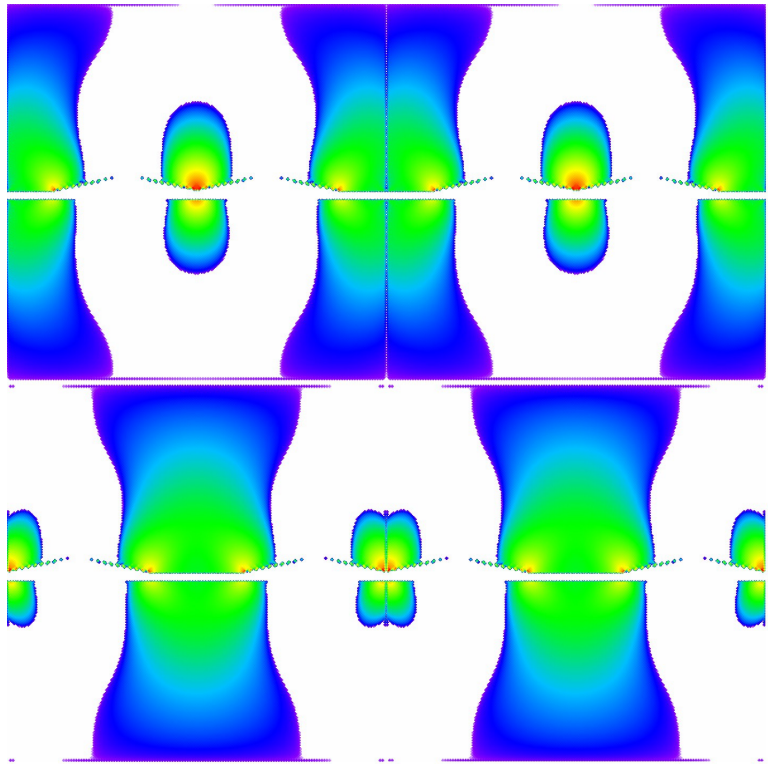

**Figure 15.** The distribution of the electron potential energy in silicon for $\Delta^{100}$ valley for the structure, periodic along the axis *x*, obtained by continuation of Fig. 13 and Fig. 14. The distances between the centers of the pyramids equal to 110 and 190 atomic layers.

Fig. 16 and Fig. 17 show the horizontal sections of the potential surfaces (in a plane (x, y)) in the case where the distance between the centers of the adjacent pyramids is 110 atomic layers.

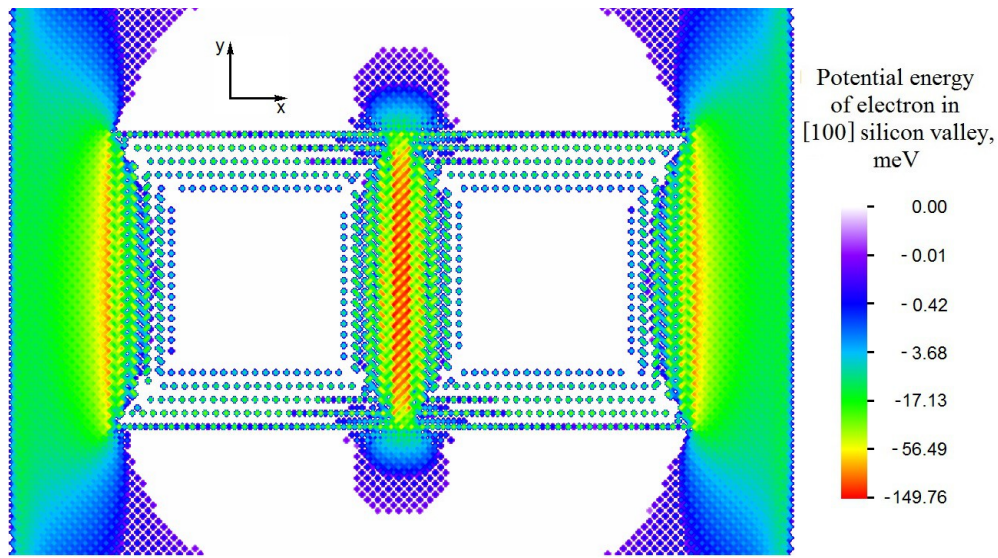

**Figure 16.** The distribution of the electron potential energy in silicon for $\Delta^{100}$ valley in the section *z* = const (3rd atomic layer of pyramids) for two pyramids with the distance between the centers of the pyramids equals to 110 atomic layers.

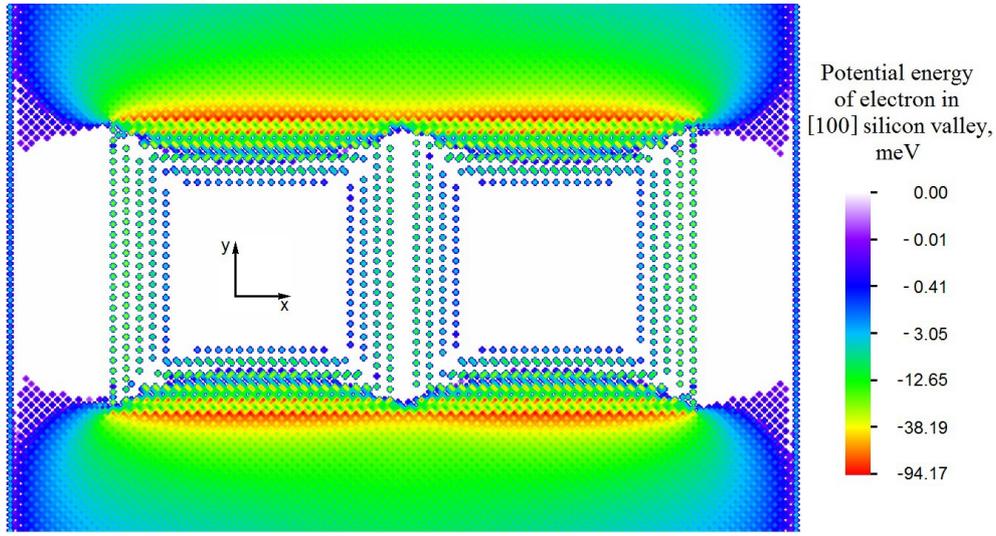

**Figure 17.** Distribution of the electron potential energy in silicon for $\Delta^{010}$ valley in the section $z$ = const (3rd atomic layer of pyramids) for two pyramids with the distance between the centers of the pyramids equals to 110 atomic layers.

The values of the electron potential energy for the main calculated potential wells are collected for comparison in the table.

| Valley | Well | The distance between the centers of pyramids (atomic layers) | | | | |
|---|---|---|---|---|---|---|
| | | ∞ | 190 | 170 | 130 | 110 |
| $\Delta^{001}$ | upper | -327.03 | -313.07 | -315.97 | -316.37 | -315.50 |
| $\Delta^{001}$ | lower | -151.71 | -144.92 | -146.38 | -147.01 | -145.69 |
| $\Delta^{100}$ | upper | -109.18 | -150.27 | -124.51 | -129.27 | -149.76 |
| $\Delta^{100}$ | lower | -46.74 | -70.82 | -70.33 | -72.43 | -82.73 |
| $\Delta^{010}$ | upper | -109.18 | -109.88 | -109.53 | -108.90 | -108.63 |
| $\Delta^{010}$ | lower | -46.74 | -47.48 | -47.12 | -46.55 | -46.40 |

**Table.** The minimal values of the electron potential energy (meV) for the main calculated potential wells at different distances between two pyramids.

For considered clusters, we deal with several potential wells of the same type, localized near different pyramids. The values of energies for the most deep of such wells of one type are given in the table. Therefore, for example, for $\Delta^{100}$ valley, for two types of potential wells (upper and lower) in the case of the distance between the centers of the pyramids in 190 atomic layers there are shown in the table the values of energies not for central wells (between the pyramids), but only for potential wells near the cluster boundaries. However, pyramids interact with pyramids outside the cluster near the cluster boundary according to the (almost) periodic boundary conditions. Therefore, the values for the upper wells -149.76 meV and -150.27 meV (for distances of 110 and 190 atomic layers), in fact, relate to the same well, but in the first case the well is at the center of the cluster, and in the second case it is at the cluster boundary. Practical coincidence of these two values once again confirms the proximity of the used boundary conditions to periodic conditions.

The difference is more considerable for the lower potential wells: -70.82 meV and -82.73 meV. But this is due simply to the fact that we do not calculate the strain tensor components for atoms of the last coordination sphere. However, the minimum energy for the lower potential well for this structure just had to be achieved in atoms of the last coordination sphere under periodic conditions. Therefore, the value -70.82 meV corresponds not to a minimum of the potential well, but relates to the neighboring atoms, that is, in this case, to atoms of the penultimate coordination sphere

REFERENCES

* Electronic address: quant@ict.nsc.ru